\documentclass[pdflatex,sn-mathphys-num]{sn-jnl}

\usepackage{graphicx}%
\usepackage{multirow}%
\usepackage{amsmath,amssymb,amsfonts}%
\usepackage{amsthm}%
\usepackage{mathrsfs}%
\usepackage[title]{appendix}%
\usepackage{xcolor}%
\usepackage{textcomp}%
\usepackage{manyfoot}%
\usepackage{booktabs}%
\usepackage{algorithm}%
\usepackage{algorithmicx}%
\usepackage{algpseudocode}%
\usepackage{listings}%
\usepackage[T1]{fontenc}    
\usepackage[utf8]{inputenc} 
\usepackage{lmodern}        
\usepackage{mathrsfs} 

\theoremstyle{thmstyleone}%
\theoremstyle{thmstyletwo}%

\theoremstyle{thmstylethree}%

\begin{document}

\title[Article Title]{High resolution Fluorescence lifetime IMaging Micro-Endoscopy (FLIMME) using a single multimode fiber}


\author*[1]{\fnm{Victoria} \sur{Fay}}\email{victoria.fay@epfl.ch}

\author[1]{\fnm{Ye} \sur{Pu}}\email{ye.pu@epfl.ch}

\author[2]{\fnm{Omer} \sur{Tzang}}\email{omer.tzang@modendo-inc.com}

\author[2]{\fnm{Antonio} \sur{Caravaca}}\email{antonio.caravaca@modendo-inc.com}

\author[3]{\fnm{Rafael} \sur{Piestun}}\email{rafael.piestun@colorado.edu}

\author[4,5]{\fnm{Genrich} \sur{Tolstonog}}\email{Genrich.Tolstonog@chuv.ch}

\author[4,5]{\fnm{Christian} \sur{Simon}}\email{Christian.Simon@chuv.ch}

\author[1]{\fnm{Demetri} \sur{Psaltis}}\email{demetri.psaltis@epfl.ch}

\author[1]{\fnm{Christophe} \sur{Moser}}\email{christophe.moser@epfl.ch}

\affil[1]{\orgdiv{Laboratory of Applied Photonics Devices}, \orgname{Ecole Polytechnique Fédérale de Lausanne (EPFL)} \orgaddress{\street{}, \city{Lausanne} \postcode{1015}, \state{} \country{Switzerland}}}

\affil[2]{\orgdiv{Modendo}, \orgname{} \orgaddress{\street{1815 Bluebell Avenue}, \city{Boulder}, \state{CO}, \postcode{80302},\country{USA}}}

\affil[3]{\orgdiv{Department of Electrical, Computer and energy Engineering}, \orgname{ University of Colorado Boulder}, \orgaddress{\street{} \city{Boulder},  \state{CO},\postcode{80309}, \country{USA}}}

\affil[4]{\orgdiv{Department of Otolaryngology,Head and Neck Surgery}, \orgname{Centre Hospitalier Universitaire Vaudois (CHUV) }, \orgaddress{\street{Rue du Bugnon 46}, \city{Lausanne}, \postcode{1011} \state{}, \country{Switzerland}}}

\affil[5]{\orgdiv{AGORA Cancer Research Center}, \orgname{ } \orgaddress{\street{Rue du Bugnon 25A}, \city{Lausanne}, \postcode{1005} \state{}, \country{Switzerland}}}


\abstract{Endoscopic optical imaging using a single multimode fiber (MMF) has emerged as a promising approach for highly compact, minimally invasive, and high-resolution imaging. Unlike conventional fiber bundles, MMF-based endomicroscopes exploit the controlled excitation of multiple spatially overlapping modes in a single MMF. of core diameters of tens of micrometers. to deliver and collect light to form images with sub-micrometer resolution. Here, we introduce a fluorescence lifetime imaging microscopy (FLIM) modality to the MMF endomicroscope. We use amplitude modulation of a 405 nm single-mode light source at radio frequency (RF) and lock-in detection of autofluorescence to obtain intensity and lifetime images at sub-micrometer resolution. We experimentally demonstrate the capability of the ultrathin endomicroscope to perform label-free imaging in thick ex vivo murine submandibular gland tissue. With a temporal resolution of 0.03 ns, the FLIM images show distinguished structures of lifetime differences down to 0.5 ns. The combination of sub-micrometer fluorescence intensity and lifetime images in a minimally invasive endomicroscope opens new avenues for label-free cancer detection.  }

\keywords{fluorescence lifetime imaging (FLIM), endoscopy, ex-vivo imaging, multimode fiber (MMF), minimally-invasive}



\maketitle

\section{Introduction}\label{intro}

Fluorescence lifetime imaging (FLIM) \cite{Becker1989, Szmacinski1993, Gadella1993, Bastiaens1999} has emerged as a powerful contrast mechanism for analyzing in-vivo biological samples in a label-free manner. Endogenous fluorophores such as NADH, FAD, FMN, and extracellular matrix components like collagen and elastin contribute to both intensity and lifetime contrast, offering enhanced biochemical specificity. The combination of steady-state fluorescence intensity and lifetime measurements improves the distinction between healthy and cancerous tissues \cite{bib1,bib2,bib3,bib4}.

Unlike autofluorescence intensity alone, which is sensitive to illumination geometry, photobleaching, concentration and tissue morphology, FLIM captures temporal decay characteristics that are more robust to these confounding factors, offering improved specificity and stability, especially in complex biological environments \cite{bib12,bib13}. For instance, Manjunath et al. showed that intensity-based autofluorescence varies significantly across patients, and in some cases, malignant tumors are indistinguishable from healthy tissue \cite{bib1}. Betz et al. similarly found that tumor detection based solely on autofluorescence varied with tumor morphology and location \cite{bib7}. These limitations highlight the advantage of FLIM’s sensitivity to local molecular environments, which provides added contrast independent of intensity and can capture subtle biochemical changes related to metabolism and extracellular matrix remodeling, offering more robust tissue characterization \cite{bib4,bib8}. The clinical relevance of fluorescence lifetime in cancer diagnosis was demonstrated as early as 1997 by Wagnières et al., highlighting its value in delineating tumor margins \cite{bib4}.

Lifetime contrast is influenced by environmental factors such as pH, oxygen, ion concentrations, redox state, and molecular binding interactions, making it suitable for probing dynamic physiological processes. It has been used to investigate transient calcium signaling, membrane fluidity, and protein–ligand interactions \cite{bib12}. Clinically, FLIM has already been used in diagnosing cancers of the gastrointestinal tract, lungs, skin, head and neck, and brain, as well as in ophthalmology and cardiology \cite{bib13}.

FLIM is particularly valuable for studying metabolic states in cancer, as changes in the NADH/NAD$^+$ and free/bound NAD(P)H ratios—along with FAD redox dynamics—serve as reliable biomarkers of glycolysis and oxidative phosphorylation \cite{bib11}. This allows for delineation of tumor margins and monitoring of metabolic activity in adjacent immune or neuronal cells, extending applications to immunotherapy and cancer neuroscience.

In dermatology, FLIM can distinguish structural components like collagen and elastin, and detect pathophysiological changes in keratinocytes or melanocytes based on lifetime shifts \cite{bib10}. Traditional one-photon FLIM is limited by shallow tissue penetration. Multiphoton FLIM systems can improve penetration depth to the hundred micrometer scale and have enabled noninvasive skin sectioning, tumor delineation during brain surgery, and assessment of pharmacokinetics in vivo, making them promising alternatives to biopsy in pharmaceutical studies \cite{bib10}. However, most clinical use remains constrained to surface-accessible regions like the skin \cite{bib11}.

Coupling FLIM with endoscopy addresses this limitation by enabling access to remote sites in vivo, such as the oral cavity, larynx, and pharynx—key in head and neck cancer detection, where open surgery or bulky microscopes are impractical. Researchers have developed coarse spatial resolution endoscopic FLIM platforms using flexible or rigid probes based on multimode fibers (MMFs), ultraviolet excitation, and time-resolved detection \cite{bib13, bib14, bib15, bib17}. In these systems, the optical resolution is equal to the diameter of the fiber core (several hundreds of micrometers). For example, \cite{bib13} presents a 355 nm picosecond laser through a 365 $\mu$m core diameter MMF. The autofluorescence lifetime  from collagen (390 $\pm$ 20 nm), NADH (470 $\pm$ 14 nm), FAD (542 $\pm$ 25 nm), porphyrins (629 $\pm$ 26.5 nm)is collected by the MMF with high temporal resolution. In initial patient studies, cancerous tissue showed shorter lifetimes in NADH and FAD channels compared to normal tissue, while collagen displayed the opposite trend \cite{bib15}. Support vector machine (SVM) algorithms achieved classification accuracies above 80\% \cite{bib16}, and 94\% sensitivity \cite{bib17,bib18}, highlighting the promise of machine-learning-aided FLIM diagnostics.

While these systems provide valuable diagnostic insights, they remain limited by resolution and penetration depth. Current fiber-based FLIM endomicroscopes lack the ability to scan a diffraction-limited micrometer scale spot across tissue, meaning they cannot resolve cellular-level structures - a critical limitation when trying to distinguish fine anatomical or pathological features. For instance, tumor margin identification accuracies depend on high-resolution features. Furthermore, the fiber bundles utilized in these systems are limited by their large diameter, causing tissue damage, as core spacing is critical to prevent crosstalk between individual fibers due to evanescent fields, thus compromising resolution \cite{bib19}.

One promising route to achieve high spatial resolution imaging in deep tissue is wavefront shaping through multimode fibers. Multimode fibers have the advantage of propagating all modes through a single core. However, modes experience modal dispersion,  resulting in a speckle-like pattern at the output when illuminated by a laser source with a long coherence length \cite{bib20}. Nonetheless, the information is retained in the speckle pattern and several studies have exploited wavefront shaping to obtain high-resolution imaging through the MMF by scanning diffraction-limited spots at the distal end of the MMF and collecting fluorescence \cite{bib21,bib22}. The transmission matrix (TM) method consists of  sending known input fields through the MMF via a spatial light modulator and measuring the complex output fields as its distal end to construct a matrix corresponding to the linear transformation. With tailored wavefront generation, the TM can be used to generate almost arbitrary output fields. For example,  focused spots can be formed and digitally scanned across the fiber output \cite{bib23,bib24,bib25}. The technique has been applied  to in-vivo brain imaging. \cite{bib9,bib27}. 

In this paper, we introduce a FLIM modality to a MMF MicroEndoscope, which we call FLIMME. FLIM has been implemented using approaches based on the time domain (TD-FLIM) or the frequency domain (FD-FLIM). While time-domain techniques rely on short picosecond pulsed lasers and photon counting to recover exponential decay profiles, frequency-domain methods modulate continuous-wave excitation and extract lifetimes from the phase shift between excitation and emission signals. For multimode fiber–based endoscopy, FD-FLIM is particularly advantageous because it enables the use of coherent continuous wave (cw) lasers, which avoids the severe pulse broadening effect due to the modal dispersion in MMF. We therefore use a CW laser of a long coherence length in the wavefront shaping approach.

\begin{figure}[h!]
\begin{center}
\begin{tabular}{c}
\includegraphics[width=0.75\textwidth]{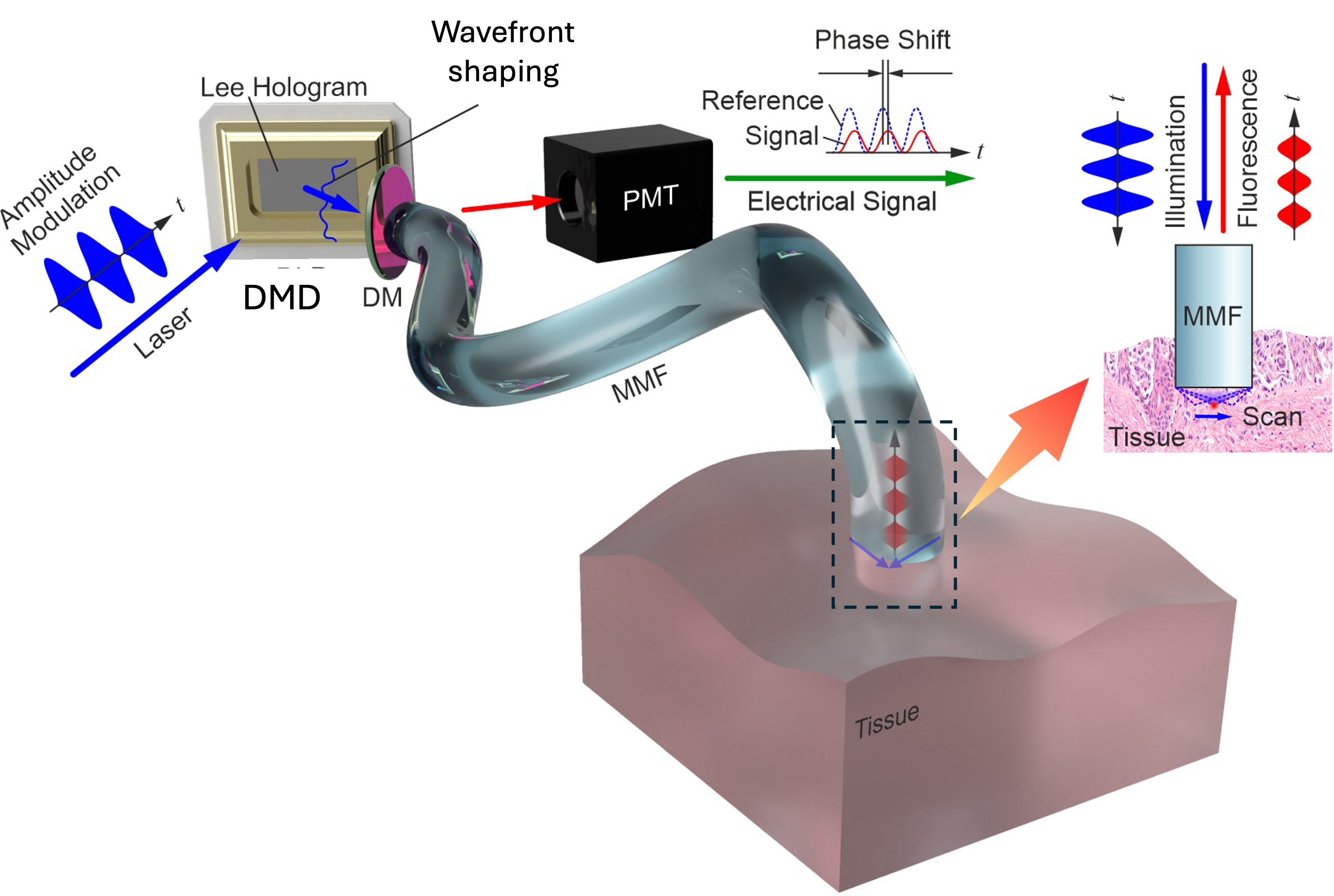}
\end{tabular}
\end{center}
\caption 
{ \label{fig0}
High-resolution FLIM endoscopy using a single MMF: The wavefront of an amplitude-modulated near-UV laser at high frequency is shaped to obtain a tight, diffraction-limited focal spot at the output of a MMF. Endogenous fluorescence is emitted from the focus spot with a small delay as a function of its lifetime compared with the modulated excitation light. The photomultiplier detects the fluorescence signal returning from the MMF, whose phase shift from the modulation reference, and therefore the fluorescence lifetime, is extracted. MMF: multimode fiber. DMD: Digital Micromirror Device. DM: dichroic mirror. PMT: photomultiplier.} 
\end{figure}

Building on recent progress in controlling light through multimode fibers, we demonstrate here a wavefront-shaping MMF FLIMME providing minimally invasive and high-resolution fluorescence lifetime imaging of endogenous fluorophores (Fig. \ref{fig0}). This approach combines the metabolic and structural sensitivity of lifetime contrast with the spatial resolution of MMF based wavefront shaping, opening opportunities for applications such as, for example, intraoperative cancer margin detection in head and neck surgery.

\textcolor{red}{}

\section{Methods}\label{sec11}

\bmhead{Setup}  

Excitation wavelengths in the UV range are suitable for exciting relevant endogenous fluorophores. However, these wavelengths can damage DNA \cite{Klak2020}, making them unsuitable for surgical applications beyond imaging excised tissues or tissues intended for excision. Near-UV wavelengths, such as 405 nm, have previously been shown to excite NADH \cite{bib5}, despite lying at the edge of its excitation spectrum and resulting in approximately a tenfold decrease in efficiency compared to excitation at 355 nm \cite{bib6,bib7}.

TD-FLIM is commonly employed in commercial microscopes, where short laser pulses excite the sample and photon arrival times are recorded using time-correlated single photon counting (TCSPC). Because only one photon is typically detected per excitation cycle, many repetitions are needed to build statistically reliable histograms. High photon rates can introduce pile-up artifacts, where early-arriving photons dominate due to detector dead time. Moreover, pulsed lasers suffer from pulse broadening when propagating through MMFs which render ultrafast pulsed sources at 405 nm ill-suited for MMF-based endoscopy.

In contrast, frequency-domain FD-FLIM enables lifetime measurements using amplitude-modulated continuous-wave (CW) lasers, which is suitable for coherent wavefront shaping in MMFs. FD-FLIM theory is well understood. The fluorescent sample is first excited by a light source modulated at a specific frequency, typically  in the tens of Megahertz range. The sample then emits fluorescence at the same frequency, but with a phase delay and reduced amplitude relative to the excitation. The fluorescence lifetime $\tau$ is calculated from the phase delay:

\begin{equation}
    \tau=\frac{1}{2\pi f}tan(\phi),
\end{equation} where $f$ is the modulation frequency and $\phi$ is the phase shift between the excitation reference and the detected fluorescence. For fluorescence exhibiting multi-exponential decays, measurements at multiple modulation frequencies can be combined to extract multiple lifetimes. In this study we use one modulation frequency which yields a single fluorescence lifetime, however the method can be extended to multiple modulation frequencies yielding multiple fluorescence lifetimes.

For FLIMME, amplitude modulation is imparted to all the spatial modes propagating in the fiber. The different propagation modes travelling at slightly different speeds and interfering at the output might introduce a distortion of the time modulated focus spot \cite{bib29}. The maximum temporal spread between the highest and lowest order modes can be estimated in a step index fiber as:

\begin{equation}
    \Delta t = \frac{n_{core}L}{c}(1-\sqrt{1-NA^2}), 
\end{equation} where $n_\mathrm{core}$ is the refractive index of the fiber core, $L$ is the fiber length, NA is the fiber numerical aperture, and $c$ is the speed of light in vacuum. With a MMF length of 5 cm, the intermodal temporal spread is on the order of a few picoseconds, which is negligible compared with the nanosecond timescale of modulation. We chose a modulation frequency of 80 MHz to measure lifetimes up to 8 nanoseconds. Given the modulation period of 12.5 ns, it is reasonable to assume that the modulated focus emerging from the MMF preserves the input modulation with high fidelity; thus, a sinusoidal modulation at the input remains essentially a sine wave at the output.

i.e. a sinusoidal modulation at the MMF input remains sinusoidal at the MMF output.

Blue violet Laser diodes suitable for FD-FLIM can be internally modulated at MHz frequencies, although, at high output power, the amplitude modulation creates instabilities in the longitudinal modes. Electro-optic external modulation, in contrast, allows us to modulate the amplitude without changing the internal lasing characteristics of the laser diode. In our system, a polarized 405 nm CW laser (Topmode, Toptica) with 50 meters coherence length was externally modulated at 80 MHz using a resonant electro-optic modulator (EO-AM-R-80, Thorlabs). The high-extinction-ratio Glan–Thompson polarizer (Thorlabs) placed at the output of the EO modulator lets only one linear polarization through, thus creating an amplitude modulation. 

A commercially available, scanning, minimally invasive endomicroscope (Neurovu, Modendo, Boulder, CO, USA) was modified to enable the 405 nm light source to perform FLIM imaging. The modulated excitation laser is delivered to the system by coupling it to a single mode fiber. At the output of the single mode fiber, there is approximately 10 mW average power.

The system employs a digital micromirror device (DMD; V-7001, Vialux) operating at maximum speed (22.727 kHz) illuminated by the 405 nm light from the single-mode fiber.  The calibration of the endomicroscope is performed via the transmission matrix (TM) method. Once the matrix is measured, the phase patterns creating the focus spots within the output field of view of the MMF \cite{bib8,bib9} are computed using the matrix. The holographic Lee method is used to display the phase patterns on the binary DMD. 

\begin{figure}[h!]
\begin{center}
\begin{tabular}{c}
\includegraphics[width=\textwidth]{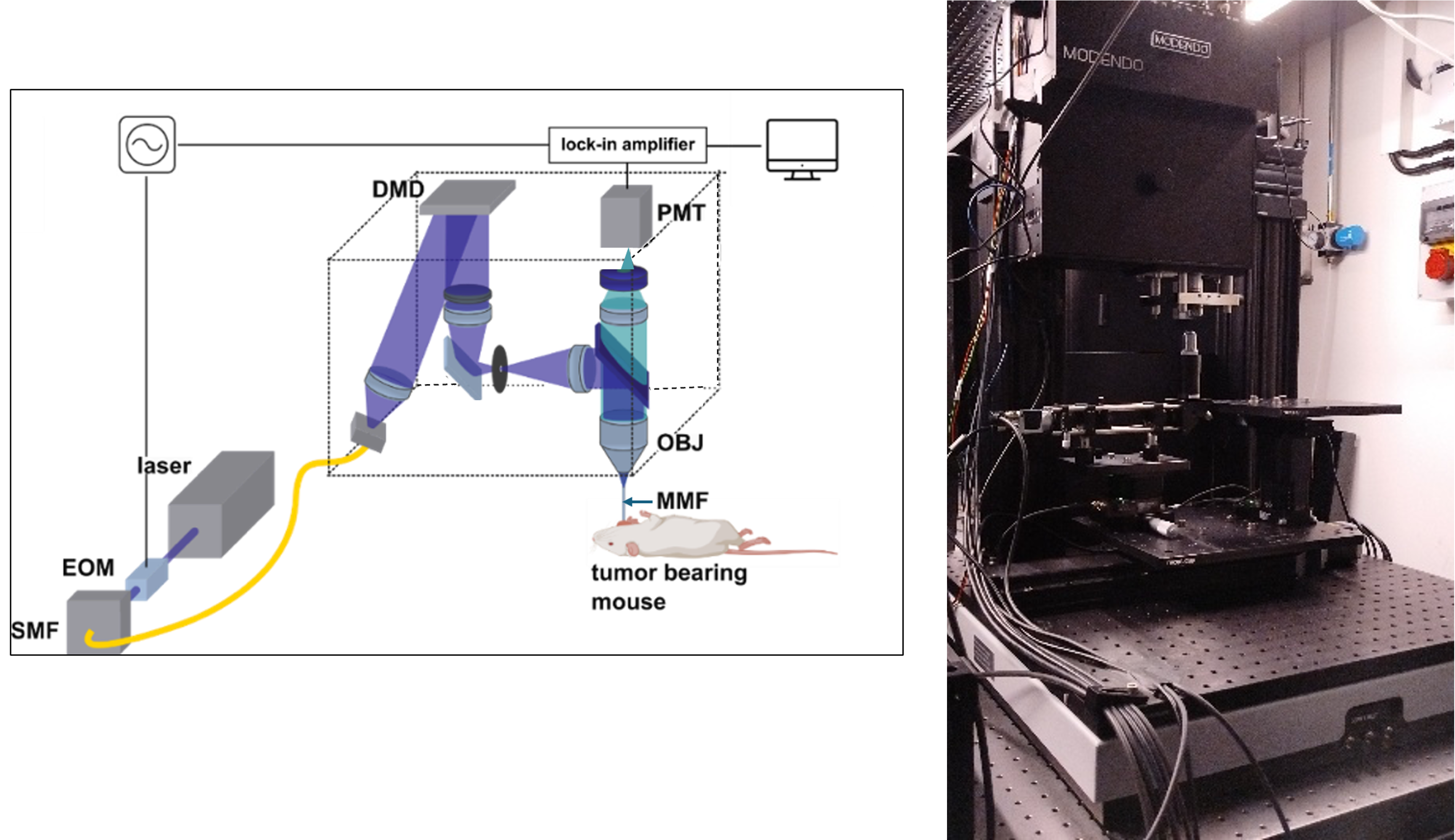}
\end{tabular}
\end{center}
\caption 
{ \label{fig:setup}
Minimally invasive endomicroscope: (a) Schematic of the optical setup. A 405 nm laser is externally modulated at 80 MHz using an electro-optic modulator and delivered to the system via a single-mode fiber. After wavefront shaping with a digital micromirror device, the light is coupled into the proximal side of the multimode fiber (MMF) and then focused at its distal side. Fluorescence collected through the same fiber is spectrally filtered to match the spectral bands of NADH and FAD and detected by a photomultiplier tube (PMT). (b) Tabletop MMF FLIMME system} 
\end{figure}

Before imaging, the 3D stage onto which the sample is positioned allows for precise positioning of the sample laterally and vertically towards the MMF probe. During scanning, the fluorescence from the sample is collected through the same MMF and directed through a choice of two optical bandpass filters corresponding to NADH (466/40 nm, Semrock) and FAD (540/50 nm, Semrock). A motorized flipper mount (8892-K-M, Newport) selects one filter at a time. The Photomultiplier tube (PMT,H11901-110, Hamamatsu) detects the returning fluorescence and its electrical output was amplified by a transimpedance amplifier (TIA), which is then demodulated with homodyne detection to extract the amplitude and phase of the signal relative to the modulation signal as the reference. A compact custom analog lock-in amplifier (LIA) was built for the demodulation. A schematic and photos of the setup is shown in Fig. \ref{fig:setup}. This FLIM implementation provides high signal-to-noise ratios, is compatible with wavefront shaping through MMFs, and provides label-free FLIM imaging of NADH and FAD with microscopic resolution in an ultrathin endomicroscope.

\bmhead{Sample preparations} 

Liquid solutions of Coumarin dyes, which have known fluorescence lifetimes, were prepared to calibrate the temporal delay of the FD-FLIM system. Concentrations of 100 $\mathrm{\mu}$M and 10 $\mathrm{\mu}$M of Coumarin 102 (C102) and Coumarin 334 (C334) in DMSO were used to perform calibrations at two PMT gain settings.

Agarose gels (UltraPure Agarose, Invitrogen) at 1\% in phosphate-buffered solution (pH 7.4) were used as a matrix to host microspheres.  For each sample, 20 $\mathrm{\mu}$L of 10× diluted Yellow-Green (YG) microspheres (2$\mathrm{\mu}$m diameter; Fluoresbrite 18338, Polysciences) and 200 $\mu$L of 100 $\mathrm{\mu M}$ Coumarin dye were added to 2 mL of the liquid agarose prior to gelation. The known microspheres' size and fluorescence lifetimes are used to to assess the spatial resolution and temporal accuracy of the FLIM endomicroscope system.

The true fluorescence lifetimes of all samples were measured on the same day using a commercial TD-FLIM microscope (LEICA SP8) with 440 nm excitation. For agarose gels, the volume of interest was sliced and placed on a coverslip for imaging with both the endoscopic system and the microscope.

Next, the endomicroscope is tested using ex-vivo mice samples. All mice experiments were approved by the Ethics Committee for Animal Experimentation of Switzerland and performed following institutional guidelines. Eight week old female C57BL/6JRj mice (JANVIER LABS, Le Genest-Saint-Isle, France) were used for fresh organ collection.Mice were placed under 4.5\% isoflurane anesthesia for induction, then humanely sacrificed under cervical dislocation. Organ collection was then performed directly after sacrifice and placed in 0.9\% NaCl. In case of immediate imaging, samples were kept on ice and in case of imaging at a later stage, samples were placed in 90\% FBS and 10\% DMSO then placed in a -80 degrees celsius freezer in a slow freezing box. The samples were then thawed at 37 degrees and immersed in 1X PBS solution during imaging.




\bmhead{FD-FLIM delay calibration} 
Optical elements in the system between modulation and detection introduce an optical delay, and cables contribute an electronic delay. Calibration of the FD-FLIM system was performed using liquid dye solutions of known lifetime. The liquid solution  provides a stable uniform lifetime distribution thoroughout the volume.The phase delay of the system was calculated using the C102 reference sample (lifetime $\mathrm{\tau}$=3.22 ns) according to:

\begin{equation}
    \phi_{delay}=\phi_{measured}-\mathrm{arctan}(2\pi f \tau).
\end{equation} 

where the $\mathrm{arctan}$ term is the phase shift caused by the fluorescence lifetime $\mathrm{\tau}$ and $\mathrm{\phi_{measured}}$ is the measured phase shift provided by the Lockin amplifier. Measured lifetimes for subsequent measurements were corrected by subtracting this phase delay.

To assess the effects of wavefront shaping and focal spot scanning, the dye solutions were imaged to verify the homogeneity of the measured lifetime distribution.

\bmhead{Imaging} 

Prior to each imaging session, the fiber tip is inspected and either cleaned or replaced. The transmission matrix is then measured using an orthogonal set of $128^2$ Hadamard patterns. This choice provides optimal mode coverage within the fiber and provides the best focal spot enhancement. For fibers with a 100 $\mathrm{\mu}$m core, the field of view (FOV) is set to 160 × 160 pixels for full-resolution calibration (corresponding to a FOV of 32 × 32 $\mathrm{\mu}$m) or 144 × 144 pixels for binned pixels on the calibration camera (FOV = 57.6 $\mathrm{\mu}$m). The calibration plane is slightly offset below the distal tip of the fiber to ensure focusing just outside the fiber. The calibration procedure takes approximately 7 minutes for the binned configuration and remains stable throughout the subsequent imaging measurements.

For imaging, the sample stage is positioned beneath the fiber. The sample is then raised slowly until contact with the MMF probe, indicated by a sharp increase in fluorescence intensity. This position is defined as z = 0. The fiber is then advanced in 5 $\mathrm{\mu}$m increments. The focal spot is then scanned at the maximum refresh rate of the DMD (22.727 kHz). Acquisition begins 10 $\mathrm{\mu}$s after detection of the falling edge of the DMD trigger, allowing the lock-in amplifier to average over 1,000 modulation periods during the 12.5 $\mathrm{\mu}$s dwelling time at each spot. Note that this large number of modulation periods validates the expression for the phase shift in equation (1), which requires that the measurement time (12.5 $\mathrm{\mu}$s) is much larger than the fluorescence lifetime (few $\mathrm{\ ns}$).

The fiber was cleaned between each insertion using a bath of isopropanol and lens tissue.

\section{Results}\label{sec2}


\begin{figure}[h!]
  \centering
  \includegraphics[width=\textwidth]{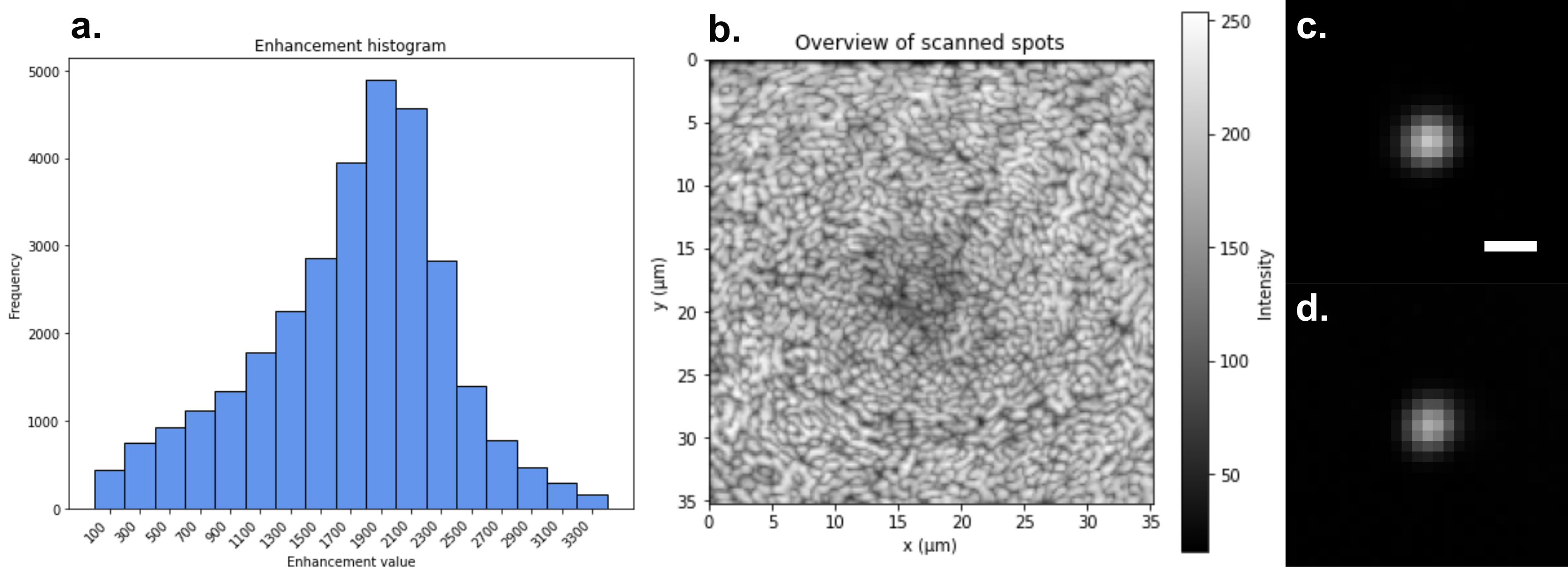}
  \caption{\label{fig1}
  Focusing performance of the endomicroscope: (a) Histogram of focal spots enhancement in a
  $1225~\mu\mathrm{m}^2$ area using a $30~\mathrm{mm}$ long MMF.
  (b) 2D maps of focal-spot intensities (0–255; 8-bit) showing less-intense spots in the center
  (lower enhancement). (c,d) Example of two focal spots illustrating a round and slightly
  elliptical shape. Scale bar: $1~\mu\mathrm{m}$.}
\end{figure}

The optical performance of the system is first evaluated in terms of spatial focusing capabilities and temporal fluorescence lifetime resolution. The endomicroscope is then validated by demonstrating ex-vivo deep tissue imaging using only endogenous fluorescence. Focusing enhancement, defined as the ratio of the intensity at the focus spot to the average background intensity across the fiber core, reaches an average value of 2000 for 30 mm long fibers with a 100 $\mathrm{\mu}$m core (see Fig.\ref{fig1}a)), while values of 3000 are reached using shorter fibers. To achieve these high enhancements, 128 × 128 Hadamard patterns are used as the set of input patterns for the transmission matrix (TM) calibration. The set of Hadamard patterns span the entire DMD array, resulting in a Fourier plane size slightly larger than the fiber core. Thus, to excite higher-order modes in the fiber, the patterns are deliberately laterally shifted with respect to the fiber core. The gain in enhancement is accompanied with spatial inhomogeneity of the focal spot intensities: the central focal spots have slightly lower enhancement than the focal spots at the edge (see Fig.\ref{fig1}b). The average focal spot size, measured as FWHM, is 0.716 $\mathrm{\mu}$m $\pm $0.055 $\mathrm{\mu}$m, with a mean ellipticity of 7.2\%. The obtained spot size is slightly larger than expected (0.716 vs 0.573 $\mathrm{\mu}$m)  given the numerical aperture of the fiber (NA=0.36 $\mathrm{\mu}$m). We attribute it to the non perfect control of the higher order modes in the fiber, which reduces the effective numerical aperture from 0.36 to 0.29. For these measurements,  the spots are scanned with a  0.2 $\mathrm{\mu}$m interval. An example of typical focal spots shape is illustrated in Fig.\ref{fig1}c.

\begin{figure}[h!]
\begin{center}
\begin{tabular}{c}
\includegraphics[width=0.65\textwidth]{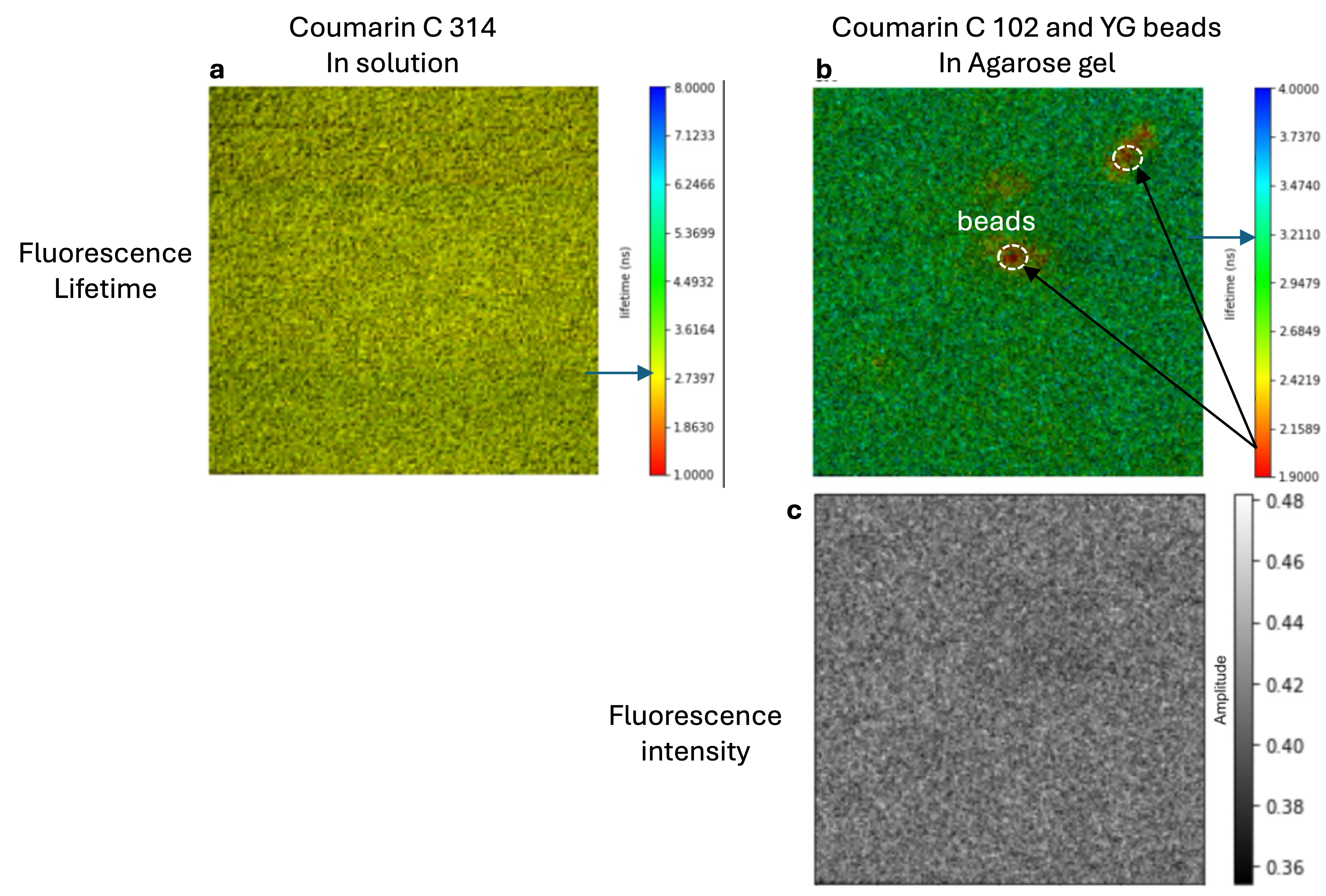}
\end{tabular}
\end{center}
\caption 
{ \label{fig3} FLIM images after calibration: (a) Coumarin C314 in solution. The average lifetime is measured at 2.89 ns, close to the value of 2.92 ns obtained with a commercial FLIM microscope on the same sample  (b) Coumarin C102 dye in agarose gel. The average lifetime is 3.20 ns, close to the 3.22 ns value measured with the commercial FLIM microscope. Out of focus YG beads can be seen in the FLIM image but not in the fluorescence intensity image (c).} 
\end{figure} 


The FLIM temporal accuracy was assessed using dyes of known lifetime.  The amplitude  and phase (from which lifetime is extracted)  are computed from the quadrature and in-phase signal outputs of the LIA. The images are then combined to produce a FLIM image. The calibration is performed with the coumarin dye C102 in solution. As explained in the method section, the delay from the optical and electrical component is computed. This uniform delay is then subtracted, pixel by pixel, from subsequent FLIM images.  
After calibration, the lifetime of the coumarin dye C314 in solution was measured as 2.89 ns  (Fig. \ref{fig3}a) with the endomicroscope, compared to 2.92 ns with a commercial TD-FLIM microscope, showing excellent agreement. Similarly, the lifetime of C102 embedded in agarose gel was measured as 3.20 ns with the endomicroscope, compared to 3.22 ns with the commercial microscope. The suspended out of focus YG microspheres can be seen in the FLIM image, as they have a different lifetime (1.99 ns) than the background. Lifetime imaging also revealed contrast not visible in the amplitude images (Fig. \ref{fig3}c). Out-of-focus spheres exhibits an apparent lifetime 0.4 ns longer than their true value due to reduced intensity and signal-to-noise ratio, a phenomenon also observed with the commercial TD-FLIM system, where out-of-focus spheres showed increased lifetimes of 0.2 ns. These results indicate a temporal accuracy on the order of 0.03 ns for the dyes and demonstrate the system’s sensitivity to lifetime differences.

\begin{figure}[h!]
\begin{center}
\begin{tabular}{c}
\includegraphics[width=0.9\textwidth]{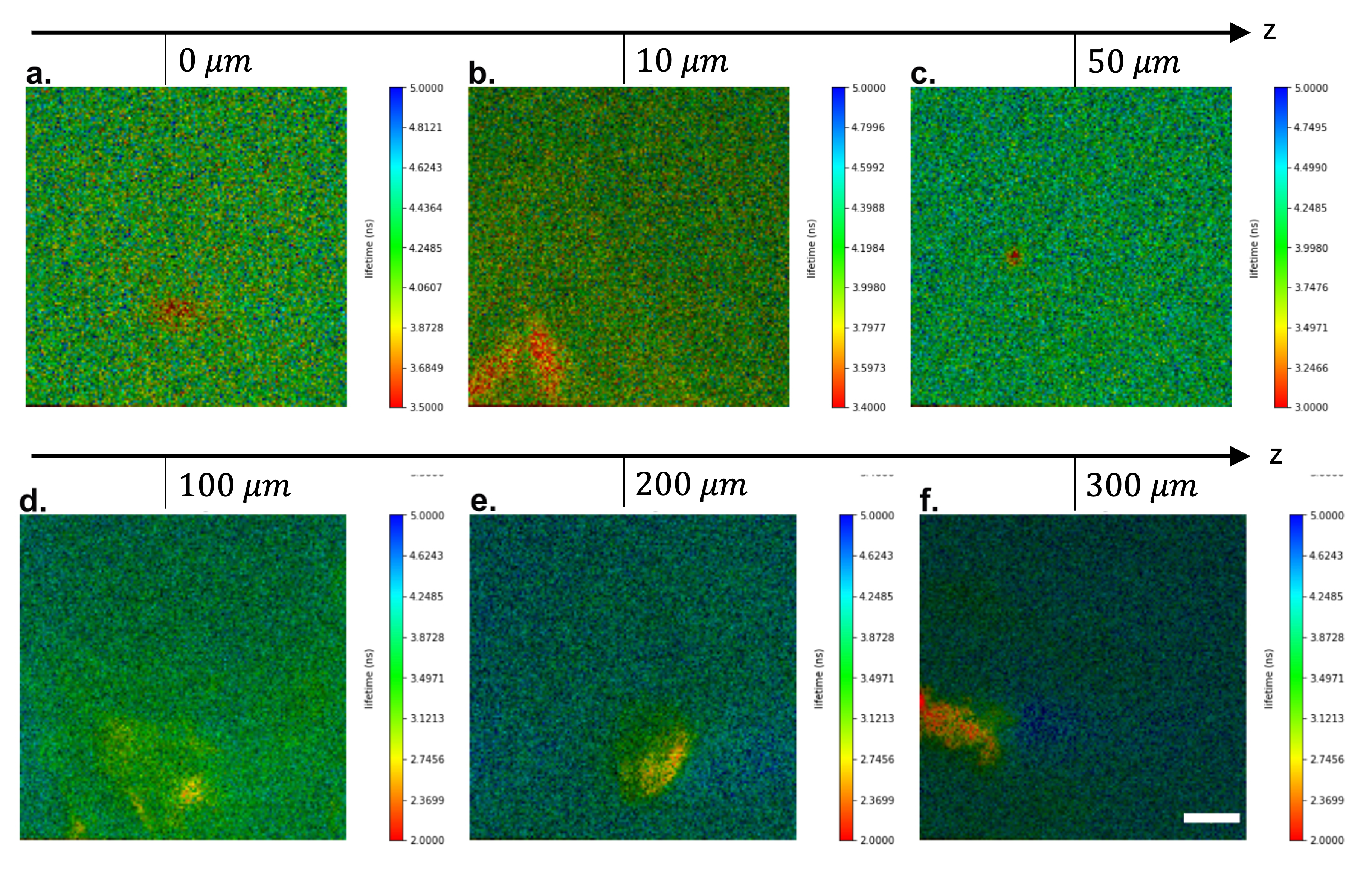}
\end{tabular}
\end{center}
\caption 
{ \label{fig4}
Ex-vivo endoscopic endogenous FLIM images of ex-vivo murine submandibular gland tissue at different depth z .} 
\end{figure} 

The endomicroscope was then tested by inserting it in an ex-vivo thick murine submandibular gland tissue. Fig \ref{fig4} shows FLIM images from endogenous fluorescence displaying microscopic contrast structures with lifetimes ranging from 2 ns to 5 ns. The structures in Fig. \ref{fig4}d,e,f have lifetime contrast of at least 0.5 ns. These images demonstrate that  the FLIMME's system’s temporal fluorescence resolution is suitable for label free imaging deep inside the tissue.

\section{Discussion}\label{sec12}

We established that modulating the laser intensity in the tens of MHz range did not interfere with the spot focusing ability.    High enhancement factors of 2000 are achieved across the multimode fiber field of view at a wavelength of 405 nm. The measured mean scanning focal spot size of 0.716 $\mathrm{\mu}$m with low ellipticity demonstrates diffraction-limited resolution, confirming that the system has the potential to resolve subcellular features. The 128 × 128 Hadamard calibration patterns, combined with wavefront shaping, enabled efficient excitation of higher-order modes, though central focal spots were slightly less efficient than those at the edges. This study demonstrated the addition of a FLIM modality to a scanning ultra-thin MMF endomicroscope via a frequency modulated approach. Importantly, the latter has the advantage of producing FLIM images at the maximum scanning speed of the DMD (22.727 KHz). The method can be adapted to any excitation wavelength. By introducing more than one modulation frequency, multiple decay times could be probed in the future.

The high spatial resolution and low signal-to-noise ratio achieved by the endoscopic system enabled precise visualization of 2 $\mathrm{\mu}$m Yellow-Green microspheres, demonstrating the capability to resolve small fluorescent features. Measurements on liquid dye solutions and agarose-embedded microspheres indicate excellent agreement between the endoscopic FD-FLIM system and a commercial TD-FLIM microscope, validating the delay calibration procedure. Differences in measured lifetimes, including a 0.4 ns increase for out-of-focus microspheres, highlight the sensitivity of lifetime retrieval to both signal-to-noise ratio and precise focal positioning. From comparisons with reference lifetimes, the system achieves an estimated temporal accuracy of $\pm$0.03 ns, sufficient for quantitative biological imaging.

Endogenous imaging in thick ex vivo murine submandibular gland tissue demonstrated the ability to discriminate structures having  lifetime differences of at least 0.5 ns. This temporal resolution is adequate to resolve biologically relevant variations in NADH and FAD lifetimes, supporting potential applications in monitoring cellular metabolism or assessing tissue heterogeneity. The observed sensitivity to out-of-focus signals also emphasizes the importance of maintaining precise focal positioning when imaging layered or scattering tissues.

By combining FD-FLIM with wavefront shaping through multimode fibers, the system overcomes the trade-off between minimally invasive imaging and high spatial resolution inherent to conventional endomicroscopes. This approach enables quantitative label free lifetime imaging in anatomically challenging regions, such as head-and-neck tissue for example, while preserving metabolic contrast. The ability to detect subtle differences in endogenous lifetimes suggests applications in tumor margin delineation, monitoring of metabolic activity, and potentially in the evaluation of therapeutic response in situ.

Limitations of the current system include lack of depth sectioning. The focus spot can excite out-of-focus fluorophores and potentially skew the measured lifetimes. Spatial sectioning by digital confocal filtering has been demonstrated with a MMF microendoscope and could be applied to FLIMME \cite{Loterie2015}. Future improvements could incorporate adaptive scanning strategies to homogenize the illumination across the field of view.

\section{Conclusion}\label{sec13}

We developed a minimally invasive, wavefront-shaped FD-FLIM endomicroscope capable of high-resolution fluorescence lifetime imaging through multimode fibers. The system, termed FLIMME, achieves a sub-micrometer spatial resolution and a temporal resolution of tens of picoseconds, allowing for reliable detection of endogenous fluorophores such as NADH and FAD. Calibration against reference dyes demonstrated excellent agreement with commercial TD-FLIM microscopes, and ex vivo imaging of murine tissues confirmed the ability to discriminate lifetime differences as small as 0.5 ns.

This work demonstrates the potential of FLIMME for quantitative biological imaging in deep or surgically inaccessible tissues, offering a label-free approach to monitor cellular metabolism and tissue heterogeneity. Future improvements, including adaptive scanning to homogenize illumination and depth sectioning could further enhance imaging performance and broaden clinical applicability in areas such as cancer margin detection, immunotherapy monitoring, and functional tissue characterization.

\backmatter

\bmhead{Supplementary information}

\bmhead{Acknowledgements}
We thank the personnel at the Bioimaging and Optics platform (BIOP) at EPFL for their help on the commercial FLIM microscope

\section*{Declarations}

\begin{itemize}
\item Funding:
\begin{enumerate}
  \item TANDEM, Cancer Research Foundation grant ``Real-time Minimally Invasive Multi-modal Endoscopy (REAL-MIME)''.
  \item U.S. National Science Foundation SBIR awards \#2212906 and \#2415645.
  \item Venture Partners at CU Boulder and the Colorado Office of Economic Development and International Trade (OEDIT) awards \#DO~2020-2464 and \#DO~2022-2458.
\end{enumerate}
\item Conflict of interest/Competing interests: RP, CM and DP have a financial interest in Modendo Inc.
\item Ethics approval: All mice experiments were approved by the Ethics Committee for Animal Experimentation of Switzerland and performed following institutional guidelines. 
\end{itemize}

\noindent

\bigskip





\begin{appendices}





\end{appendices}


\bibliography{sn-bibliography}

\end{document}